\documentstyle[prd,aps,floats]{revtex}
\tighten

\begin{document}
\draft
\preprint{} 
\title{ Thoughts on galactic magnetism \thanks{%
It is a pleasure to dedicate this article to
Prof. Jayant Narlikar, my first research Guru, though he
may want to have little to do with toroidal
and poloidal fields and the infamous 'pomega' !}}

\author{ Kandaswamy Subramanian \\
National Centre for Radio Astrophysics, \\ 
Tata Institute of Fundamental Research,\\
Poona University Campus, Ganeshkind,\\
Pune 411007, India}
\maketitle

\begin{abstract}
Magnetic fields correlated on several kiloparsec scales are seen 
in spiral galaxies. Their origin could be due to amplification 
of a small seed field by  a turbulent galactic dynamo. 
We critically review the current status of the galactic dynamo, 
especially some of its problems and possible solutions. 
We also comment on the nature of seed magnetic fields,
needed to prime the dynamo.
\end{abstract}

\maketitle
\date{\today}
\section{Introduction }

Magnetic fields in spiral galaxies have strengths of order few $10^{-6} G$,
and are coherent on scales of several kpc \cite{beck96}. 
In several disk galaxies, like M51 and NGC 6946, they are
also highly correlated (or anti-correlated) 
with the optical spiral arms.
How do such ordered, large-scale fields arise?
One possibility is dynamo amplification of a weak but nonzero seed field 
$\sim 10^{-19} - 10^{-23} G$,
provided the galactic dynamo can operate efficiently to exponentiate the
field by a factor $\sim 30 - 40$.
We critically review here some of the issues 
relevant to the operation of the 
galactic dynamo, the problems which arise and possible solutions.           
We also touch upon the origin of
the seed magnetic field, needed for dynamo amplification.

The evolution of the magnetic field, in the MHD approximation, 
is described by the induction equation
\begin{equation}
{\partial {\bf B} \over \partial t} =
{\bf \nabla } \times ( {\bf v} \times {\bf B} - 
 \eta {\bf \nabla } \times {\bf B}), 
\label{induc}
\end{equation} 
provided one assumes the usual form of Ohms law  and 
neglects the displacement current term in Maxwells equation.
Here ${\bf B}$ is the magnetic field,  
${\bf v}$ the velocity of the fluid and
$\eta$ the resistivity. If $\eta \to 0$ the magnetic flux through
any area in the fluid is conserved during the
motion of the fluid. The presence of a finite resistivity allows
for a violation of such 'flux freezing' and the magnetic Reynolds
number (MRN) $R_m = vL/\eta$ measures the relative 
importance of flux freezing versus resistive diffusion. 
Here $v$ and $L$ are typical velocity and length scales of
the fluid motions. In most astrophysical contexts 
flux freezing greatly dominates over diffusion with $R_m >> 1$.

${\bf B} =0$ is a perfectly valid solution of the induction 
equation. So there would be no magnetic field generated if
one were to start with a zero magnetic field.
The universe probably did not start with 
an initial magnetic field. One therefore needs
some way of violating the induction equation and produce a cosmic
battery effect, to drive curents from a state with initially
no current. There are a number of such battery mechanisms
which have been suggested \cite{batt}.
All of them lead to only small fields, much smaller
than the galactic fields. Therefore dynamo action, due to a
velocity field acting to exponentiate 
small seed fields efficiently, 
is needed to explain observed fields.
We first briefly comment on a cosmic battery  
before discussing dynamos in detail.

\section{ Cosmic batteries and seed fields for the dynamo }

The basic problem any battery has to address is how to
produce finite currents from zero currents?
Most astrophysical mechanisms use the fact that positively and
negatively charged particles in a charge-neutral universe, do not 
have identical properties. For example if one considered a gas
of ionised hydrogen, then the electrons have a much smaller
mass compared to protons. This means that for a given pressure gradient
of the gas the electrons tend to be accelerated much more than the ions.
This leads in general to an electric field, which couples back
positive and negative charges, of the form
${\bf E}_T =- {\bf \nabla}p_e / e n_e$, where $p_e$ and $n_e$ are
the electron pressure and number density, respectively. If such a
thermally generated electric field has a curl, then by 
Faradays law of induction a magnetic field can grow.
Taking $p_e = n_e kT$ with $T$ the electron temperature we have
${\bf \nabla} \times {\bf E}_T = (c k/  e)
( {\bf \nabla } n_e / n_e)  \times {\bf \nabla } T$.
So ${\bf E}_T$ has a curl only if the density and
temperature gradients, are not parallel to each other. 
The resulting battery effect, known as the Biermann battery,
was first proposed as a mechanism for
thermal generation of stellar magnetic fields \cite{bier}. 

The Biermann battery can also  
lead to the thermal generation
of seed fields in cosmic ionisation fronts \cite{sub94}.  
These ionisation fronts are 
produced when the first ultra violet photon sources,
like quasars, turn on to ionise the intergalactic medium (IGM).
The temperature gradient in a cosmic ionisation 
front is normal to the front.  However, a component to the
density gradient can arise in a different direction, if the ionisation
front is sweeping across arbitrarily laid down density
fluctuations, associated with protogalaxies/clusters since these
in general have no correlation to the source of the
ionising photons. The resulting thermally generated
electric field has a curl, and magnetic fields
on galactic scales can grow. They 
turn out to have a strength $B \sim 3 \times 10^{-20} G$.
This field by itself is far short of the observed microgauss strength
fields in galaxies, but it can provide a seed field,
coherent on galactic scales, for a dynamo. 
Indeed the whole of the IGM is seeded with small magnetic fields.
This seed field may infact have the right symmetry
properties for the galactic dynamo modes \cite{kbeck}.
The Biermann battery has also been invoked 
to generate seed magnetic fields in galactic
or proto-galactic environments \cite{lazkul}.

Larger seed magnetic fields than above can arise
if we combine some form of dynamo action
with the battery effect. For example, if stellar dynamos
work efficiently, and some stars blow out as supernovae,
then they can seed the interstellar medium,
with significant magnetic fields. Alternatively galactic turbulence
can itself lead to a small-scale dynamo (see below)
and provide a larger seed for 
the large-scale galactic dynamo. There have also
been attempts to invoke exotic physics in the early universe
to produce primordial magnetic fields \cite{batt}.
(Infact a primordial field in the IGM, which redshifts
to a present day value of $\sim 10^{-9} G$,
and is correlated on Mpc scales, can significantly
perturb structure formation, and cause detectable anisotropies
in the cosmic microwave background radiation
\cite{substruct}.) It is fair to say at present that most
seed field generation mechanisms fall far short
of producing large-scale correlated fields at
the micro-gauss level. One does need
some form of large-scale dynamo action.

\section{ The large-scale galactic dynamo }

Disk galaxies are differentially rotating systems.
Also the magnetic flux is to a 
large extent frozen into the fluid. So any radial
component of the magnetic field will be efficiently wound 
up and amplified to produce a toroidal component.
But this results in only a linear amplification of the field. 
To obtain the observed 
galactic fields starting from small seed fields  
one should find a way to generate the radial 
component from the toroidal one.
If this can be done, the field can grow exponentially and 
one has a dynamo.

A mechanism to produce the radial field from the toroidal 
field was invented by Parker \cite{park}.
The essential feature is to invoke the effects of cyclonic turbulence in the 
galactic gas. The interstellar medium (ISM) is assumed 
to be turbulent, due to for example the effect of supernovae randomly 
going off in different regions. In a rotating, stratified 
(in density and pressure) medium
like a disk galaxy, such turbulence becomes cyclonic and
acquires a net helicity. Helical motions of the galactic 
gas perpendicular to the disk can draw out the toroidal field 
into a loop which looks like a {\it twisted} $\Omega$.
Such a loop is connected to a current 
and because of the twist this current has 
a component parallel to the original field. If the motions
have a non-zero net helicity, then the random current components
parallel to the field, add up coherently.
A toroidal current can then result from the toroidal field. 
Hence, poloidal fields can be generated from toroidal ones. 

In quantitative terms, suppose the velocity field 
is the sum of a mean, large-scale velocity ${\bf v}_0$ 
and a turbulent, stochastic velocity ${\bf v}_T$.
The induction equation becomes a stochastic
partial differential equation. The equation for
various moments of ${\bf B}$, can be derived in
two ideal limits. First when $R_m << 1$, and the distortions
to the mean magnetic field are small, and second when
$R_m >> 1$, but the turbulence
is assumed to have a delta function (or very small)
correlation in time. For galaxies the 
latter idealisation may be more relevant.

Let us split the magnetic field ${\bf B} = {\bf B}_0 
+ \delta{\bf B}$, into a mean field $ {\bf B}_0$ and a 
fluctuating component  $\delta{\bf B}$. Here the mean is defined
either as a spatial average over scales larger than the turbulent
eddy scales or more correctly as an ensemble average.
Assume the turbulence to be isotropic, homogeneous and helical. 
The action of the turbulent velocity field ${\bf v}_T$,
on the magnetic field, the $({\bf v} \times {\bf B})$ term,
then leads to an extra contribution to the {\it mean} electric field 
of the form 
$ -c {\bf E}_0= 
\alpha {\bf B}_0 - \eta_t {\bf \nabla } \times {\bf B}_0$. 
Here 
$\alpha = -(1/3) \int < {\bf v}_T(t).({\bf \nabla} \times {\bf v}_T(s)) > ds $, 
depends on the helical part of the turbulent 
velocity correlation function, and 
$\eta_t = (1/3) \int < {\bf v}_T(t).{\bf v}_T(s) > ds $, 
called the turbulent diffusion depends on the 
non-helical part of the turbulence.
Here the angular brackets $<>$ denote
an ensemble average, over the stochastic velocity field.
The induction equation for the mean field, with the extra turbulent 
component of the mean electric field, is then given by
\begin{equation}
{\partial {\bf B}_0 \over \partial t} =
{\bf \nabla } \times \left( {\bf v}_0 \times {\bf B}_0 +\alpha {\bf B}_0 - 
 (\eta + \eta_T) {\bf \nabla } \times {\bf B}_0 \right) .
\end{equation} 
This kinematic mean-field dynamo equation, 
can have exponentially growing solutions, which 
have been studied extensively in the literature \cite{beck96}. 
While the $\alpha$-effect is crucial for regeneration
of poloidal from toroidal fields, the turbulent diffusion
turns out to be also essential for allowing changes
in the mean field flux.
Also in galaxies, differential rotation 
(the $\Omega$ effect) is dominant in producing toroidal from radial fields.
The growth rates of the galactic '$\alpha$-$\Omega$ dynamo',
are typically a few times the rotation time scales,
of order $10^9$ yr.  Modulations of $\alpha$, and $\eta_T$, due to enhanced
turbulence along spiral arms, can also lead to 
bi-symmetric large-scale fields, correlated with the optical spirals
\cite{ms91}. 

Note that the mean field has a scale
limited only by the size of the system, which
can be much bigger than the scales associated with the turbulence.
In this sense one has created order from chaos.
One may be tempted to refer to this as an inverse cascade, a term which
would suggest transfer of power from smaller to larger and
larger scales. But reality is more subtle.
All scales larger than the turbulent eddy scale 
can grow simultaneously due to the $\alpha$-effect; 
but with larger scales growing slower. So the 
effect can be thought better as a long range interaction 
between the turbulent scales, and all larger scales.

A physics comment is in order at this stage. When one considers 
the effect of turbulent fluid motions on a scalar field, like
say smoke, one only gets a mean diffusion of smoke, 
associated with the 
random walking nature of turbulent fluid motions. But for 'frozen' magnetic 
fields the induction equation has terms which not only imply 
a body transport due to the random motions (${\bf v}.{\bf \nabla} {\bf B}$),
but also a term, ${\bf B}.{\bf \nabla} {\bf v}$,
which describes the generation of magnetic fields due to
velocity shear. It is this qualitative difference between magnetic
fields and smoke that leads to an alpha effect (provided
also that the motions have a non-zero average helicity), over and
above turbulent diffusion. As we see below,
it also leads to the small-scale dynamo.
Note that both $\alpha$ and $\eta_T$, depend crucially
on the diffusive (random-walk) property of fluid motion. 
So if due to some reason (see below) the fluid motion becomes
wavelike, then the above integrals average to zero,
and the alpha effect and turbulent diffusion will
be suppressed. 

In deriving the mean-field equation,
the turbulent velocities have been assumed to be given,
and unaffected by the Lorentz forces due to the 
magnetic field; at least not until the mean large-scale 
field builds up sufficiently. However this does not turn out to
be valid due to the more rapid build up of magnetic noise 
compared to the mean field, a problem to which we now turn.

\section{The small-scale dynamo and magnetic noise }

The problem with the kinematic mean field dynamo is
that it is a myth. This is because, the same turbulence which
contributes to the $\alpha$-effect, the turbulent diffusion,
and associated growth of mean fields, also leads to a more rapid
growth of small-scale fields. 

In incompressible turbulence, 
fluid particles random-walk away from each other.
This leads to stretching of the field lines
attached to these particles, and an 
exponential increase of field strength. 
The stretching will also be accompanied
by the field being squeezed into smaller and
smaller volumes. Suppose one considers a flux tube, of length $l_t$,
cross section $A$, density $\rho$ and magnetic field $B$.
Then flux freezing implies $BA =$ constant, mass conservation
$\rho A l_t = $ constant, and incompressibility 
$\rho =$ constant. So as $l_t$ increases due to random
stretching, the the magnetic
field $B \propto l_t$ increases and the cross-section
$A \propto l_t^{-1}$ decreases.

If for the moment one ignores Lorentz forces, then the sqeezing
into small volumes, stops only when diffusive scales are reached.
Typically the field can be thought of as being in
flux ropes, curved on the eddy scale, say $L$, and a thickness
of order the diffusive scale, say $r_d$
(assuming only a single scale eddy is present). 
At this stage the energy input
into the magnetic field due to random stretching
would be comparable to the energy loss in diffusion.
This gives $vB/L \sim \eta B/r_d^2$, implying 
$r_d \sim L/R_m^{1/2}$.
What happens further (whether growth or
diffusion wins out), can only be decided by a more quantitative
treatment of the problem.

A rigorous analysis of small-scale
field dynamics, was first worked out by Kazantsev \cite{kaza},
and elaborated extensively by many authors \cite{ssd},
for the simple case, where the turbulence was assumed
to have a delta function correlation in time.
We shall mainly draw upon our own work 
in Ref. \cite{sub98} .
It turns out that, the small-scale dynamo (SSD) can operate
under fairly weak conditions; that the MRN associated
with the turbulent motions be greater than
a critical value $R_c \sim 100$.
In particular the fluctuating field
tangled on a scale $l$, can grow on the 
turn over time scale of a turbulent eddy of scale $l$, 
with a growth rate $\Gamma_l \sim v_l/l$, 
provided the MRN on that scale
$R_m(l) = v_l l/\eta > R_c $.
Here $v_l$ is the velocity associated
with eddies of scale $l$. For Kolmogorov turbulence, since
$v_l \propto l^{1/3}$, the growth rate $\Gamma_l \propto l^{-2/3}$,
and so increases with decreasing $l$.
For galactic gas, with a significant neutral component,
typically, even the eddies at
the cut-off scale of the turbulence, say $l_c$, have $R_m(l_c) >> R_c$.

The spatial structure of the small-scale dynamo
generated field is also of great interest.
For this it is useful look at the behaviour of the magnetic 
correlation function, say $w(r,t) = <\delta{\bf B}({\bf x},t).
\delta{\bf B}({\bf y},t)> $, where $r= \vert {\bf x} - {\bf y}\vert$.
Here the averaging indicated by $<>$, is a double ensemble
average over both the stochastic velocity and 
fluctuating magnetic fields. 
From ${\bf \nabla}. {\bf \delta B} = 0$, one can show
that the curve $r^2 w(r)$ should enclose zero area. 
Since $w(0) $ is neccessarily positive, as one goes to
larger $r$, there must be some $r \sim d$ say, when 
$w(r)$ becomes negative. For such $r$, 
the fluctuating field at the origin, and at a separation $d$
are pointing in 'opposite' directions on average.
This can be intepreted as saying that the field lines
on average are curved on scale $d$.

For Kolmogorov type turbulence, and if $R_m(l_c) >> R_c$, 
the fastest growing mode, has $w(r,t)$ strongly 
peaked within the diffusive scale of the
cut-off scale eddies, $r= r_d(l_c) = l_c / R_m^{1/2}(l_c)$, 
changing sign at $r \sim l_c$, and rapidly decaying for larger $r/l_c$.
One can interpret such a correlation function as implying that field
is concentrated into ropes of thickness $ r_d(l_c)$ and curved 
on scales of order $l_c$. For slower growing modes,
with growth rate $\Gamma_l \sim v_l/l$, $w(r)$ extends
upto $r \sim l$, after which it decays exponentially.
For these modes, the small-scale field can be thought of
as being concentrated in rope-like structures
with thickness of order the diffusive scale $r_d(l_c)$
and curved on a scale upto $\sim l$.
In general, at the kinematic stage, the growth rate of
irreducible higher order correlations, increase with order,
indicating that the field becomes highly intermittent in space.

Note that the small-scale dynamo due to 
even the eddy at the outer scale of galactic turbulence,
will lead to the exponetial growth of small-scale
fields on a time $ \tau = L/v \sim 10^7$ yr.
(Here we have taken $L \sim 100$ pc, and a velocity scale $v \sim 10$
km s$^{-1}$.) This time is much shorter than 
the time scale $\sim 10^9$ yr for 
mean field growth. The magnetic field 
is then rapidly dominated by the fluctuating component,
before the mean field has grown appreciably.
If the energy in the small-scale component grows to equipartition with 
the turbulent energy density, the turbulence could  become
more wavelike 'Alfv\'en' turbulence, than an eddy like
fluid turbulence. So diffusive effects like the $\alpha$ and 
$\eta_T$, would get suppressed, and mean field growth stopped.
How does the galaxy escape this predicament?

\section{ Saturation of the small-scale dynamo } 

To answer this question, it is crucial to find out how the
small-scale dynamo saturates. 
We have concentrated on the possibility that 
the small scale field continues to be intermittent in 
space, when it saturates, as it was in the kinematic
regime. That it saturates as
a 'can of worms'; with peak fields being limited by
non-linear effects to values of order
or slightly larger than equipartition fields, but with 
most of the space having much smaller fields.
Then the average energy density of
the saturated small-scale dynamo generated field,
may still be sub-equipartition, since it does not
fill the volume. And the turbulence will remain
eddy like, and preserve diffusive effects like $\alpha$ and $\eta_t$.
We have given one explicit realisation of the
above idea, in Ref. \cite{sub98},
in the case of a galaxy dominated by
neutral particles.

In partially ionised plasma, the Lorentz force
acts on ions, which are only coupled to neutrals through
collisions. This leads to a 'ambipolar' drift of
ions (and hence the field) with respect to neutrals.
With ${\bf v}$ in Eq.(\ref{induc}) replaced
by the neutral fluid velocity, the effective diffusivity
changes to $\eta_{eff} = \eta +
<\delta{\bf B}^2> / ( 6\pi \rho_i \nu_{in})$. Here $\rho_i$ is 
the ion density and $\nu_{in}$ is the neutral-ion collision 
frequency. So, as the energy density in the fluctuating field 
increases, the effective magnetic Reynolds number, for
fluid motion on any scale of the turbulence
say $ R_{ambi}(l) = v_l l/ \eta_{eff}$, decreases. 
If $R_{ambi}$ could decrease to $R_c$, this itself will
lead to a SSD saturation, but 
for conditions appropriate to galactic gas,
$R_{ambi}$ remains much larger than $R_c$.
So as the the small-scale field grows in strength, it 
continues to be concentrated into thin ropy structures, 
as in the kinematic regime. 
These flux ropes are curved on the turbulent eddy scales, while their
thickness is now set by the diffusive scale determined by the
effective ambipolar diffusion.  

Other restraining effects have to then limit the SSD.
The first of these is due to the growing magnetic tension 
associated with the curved flux ropes. 
It acts to straighten out a flux rope, at a
rate determined by equating the tension force and frictional drag.
Frictional drag also damps the magnetic energy
associated with the wrinkle in the rope. 
Further, small-scale flux loops 
can collapse and disappear, causing an {\it irreversible}
sink of magnetic energy into heat. 
These non-local effects operate on
the eddy turnover time scale, when the peak field in a flux rope,
say $B_p$, has grown to a few times the equipartion value.
Their net effect is to make
the random stretching needed for the SSD
inefficient and hence saturate the SSD.
As the field is in flux ropes which do not fill the volume.
the average energy density in the saturated small-scale field is
still sub-equipartition, and $\alpha$ and $\eta_T$ are preserved.

Note that $B_p$ has to grow to a larger and larger value,
thinner the flux rope, for inefficient random
stretching to operate. This is because tension is a 
volume force ($\propto r_d^2 l_t$) while
drag acts on the surface of the rope and is $\propto r_d l_t$.
But $B_p$ cannot grow larger
than $(8\pi P_{ext})^{1/2}$,  
where $P_{ext}$ is the total pressure in the ISM.
At the same time the thickness of flux ropes is larger, 
greater the ambipolar diffusion, or smaller the ion-density.
So the SSD saturates as above,
for a given $P_{ext}$, provided $n_i$ is less than a crictical
value $n_i^c$. In the ISM, 
if $P_{ext}$ is a factor $F$ greater than the gas 
pressure (the gas assumed to have  $T \sim 10^4$K and a density $n_n$), 
one gets $n_i^c \sim 10^{-2} cm^{-3} 
(v/10 km s^{-1})^{-3} (n_n/ cm^{-3})^{2/3}(F/2)^{7/3}$.
So for a range of 'galactic' like parameters,
this picture of small-scale dynamo saturation works.

For larger ion densities $n_i > n_i^c$,
the way the SSD saturates is not very clear.
The peak field is still limited by the external pressure.
But how the flux ropes behave in the post-kinematic
stage is yet to be rigorously worked out. It is possible that, when
one starts with weak fields, the field
is first squeezed into small volumes until
limited by magnetic pressure. Subsequently,
constructive folding of the field, may lead to fusing
and thickening of the flux rope, while destructive 
folding may lead to reconnection, and dissipation. A
phenomenological model \cite{sub99} which incorporates this
thickening of flux ropes as the field builds up,
drives the SSD into saturation, when the rope
thickness becomes of order $L/R_c^{1/2}$,
the peak field reaches equipartition levels,
but with the average energy density of order 
$R_c^{-1}$ of equipartition.

Numerical simulations
of dynamo action due to mirror-symmetric turbulence \cite{meng81}
or convection \cite{bran96} have also hinted at
a saturated state of SSD 
as described above; a magnetic field concentrated
into flux ropes, occupying a small fraction of the fluid volume,
having peak fields comparable or in excess of equipartition
value but average magnetic energy density only
about $10\%$ of the kinetic energy density.
Such simulations are however limited by the MRN 
they can achieve. There have also been MHD simulations
of the SSD in fourier space, adopting some form of closure
approximation, like the EDQNM approximation \cite{pouquet},
or the DIA \cite{chandran}. They have also indicated
that the small-scale field could saturate at sub-equipartition levels.
However, why this happens and the relation of
the fourier space, to the real space calculations
which we have emphasised, is not at present clear.

There ofcourse remains the possibility that 
the SSD in a fully ionised gas saturates only when the
energy density in the noise grows comparable to that of the turbulence. 
However, the spatial structure of the small-scale field, 
is still likely to be highly intermittent. The
large-scale dynamo action will then depend on how such a field
responds to turbulent motions, especially whether the field can
reconnect efficiently \cite{vish}. If reconnection is
efficient, then it may allow diffusive transport
to still occur, through the forest of
small-scale fields, rather like Tarzan, swinging from
one rope to another crosses the jungle!
Reconnection is an important issue, which
deserves much more discussion than we have given
(cf.\cite{lazavish}). Another important issue which is just
beginning to be addressed \cite{field}, is the
calculation of $\alpha$ (or $\eta_T$) in the presence
of significant small-scale fields.

In summary, the survival of the diffusive effects needed for
large-scale dynamo action could depend crucially on whether
the SSD generated fields can saturate at sub-equipartition
levels. We feel that this may indeed be possible,
if the noise saturates as a 'can of worms'.
Note the SSD generated field is indeed 
spatially intermittent in the kinematic regime.
So when one starts from weak fields,
kinematic evolution operates
for some time, and the small-scale field is driven to
an intermittent state. The important
question is to what extent it remains so
in the non-linear regime, when it saturates.
This SSD generated field will also provide a strong seed for
the large-scale dynamo \cite{bpoed}. Indeed, in a unified
treatment of small and large-scale dynamos,
large-scale correlations are produced, from
small-scale fields in a way analogous to
'quantum mechanical tunnelling', of the 
stationary state of the SSD \cite{sub99}.
The helical, turbulent,
$\alpha$-$\Omega$ dynamo still seems to be the best bet
for explaining large-scale galactic fields.
Ultimately the galactic dynamo is a non-linear dynamo;
but a discussion of how it operates in the final
saturated regime needs more thought. 
Our thoughts on galactic magnetism have yielded 
interesting results but much remains to be done.

\end{document}